\documentclass[a4paper,10pt,twoside]{cpc-hepnp}

\usepackage{multicol}
\usepackage{graphicx}
\usepackage{booktabs}
\usepackage{amssymb,bm,mathrsfs,bbm,amscd}
\usepackage[tbtags]{amsmath}
\usepackage{lastpage}

\begin{document}

\fancyhead[c]{\small Chinese Physics C~~~Vol. XX, No. Y (2017) 0ZZZZZ} \fancyfoot[C]{\small 0ZZZZZ-\thepage}

\footnotetext[0]{Received 21 March 2017}

\title{Shape evolutions of $^{72,74}$Kr with temperature in the covariant density functional theory
\thanks{
Supported by National Natural Science Foundation of China under Grant Nos. 11105042, 11305161, and 11505157,
the Open Fund of Key Laboratory of Time and Frequency Primary Standards, CAS,
and support from Henan Administration of Foreign Experts Affairs. }}

\author{%
      Wei ZHANG $^{1,2}$%
\quad Yi-Fei NIU $^{3;1)}$\email{nyfster@gmail.com}%
}
\maketitle

\address{%
$^1$ Henan Key Laboratory of Ion-beam Bioengineering, Zhengzhou University, Zhengzhou 450052, China\\
$^2$ Key Laboratory of Precision Navigation and Technology, National Time Service Center, Chinese Academy of Sciences, Xi¡¯an 710600, China\\
$^3$ ELI-NP, Horia Hulubei National Institute for Physics and Nuclear Engineering, 30 Reactorului Street, RO-077125, Bucharest-Magurele, Romania\\
}

\begin{abstract}

The rich phenomena of deformations in neutron-deficient krypton isotopes such as the shape evolution with neutron number and the shape coexistence attract the interests of nuclear physicists for decades.
It will be interesting to study such shape phenomena using a novel way, i.e., by thermally exciting the nucleus.
So in this work, we develop the finite temperature covariant density functional theory for axially deformed nuclei with the treatment of pairing correlations by BCS approach, and apply this approach for the study of shape evolutions in $^{72,74}$Kr with increasing temperatures.
For $^{72}$Kr, with temperature increasing,
the nucleus firstly experiences a relatively quick weakening in oblate deformation at temperature $T \sim0.9$ MeV,
and then changes from oblate to spherical at $T \sim2.1$ MeV.
For $^{74}$Kr, its global minimum locates at quadroupole deformation $\beta_2 \sim -0.14$
and abruptly changes to spherical at $T\sim 1.7$ MeV.
The proton pairing transition occurs at critical temperature 0.6 MeV following the rule $T_c =0.6 \Delta_p (0)$
where $\Delta_p(0)$ is the proton pairing gap at zero temperature.
The signatures of the above pairing transition and shape changes can
be found in the curve of the specific heat.
The single-particle level evolutions with the temperature are presented.


\end{abstract}




\footnotetext[0]{\hspace*{-3mm}\raisebox{0.3ex}{$\scriptstyle\copyright$}2017
Chinese Physical Society and the Institute of High Energy Physics
of the Chinese Academy of Sciences and the Institute
of Modern Physics of the Chinese Academy of Sciences and IOP Publishing Ltd}%


\section{Introduction}

The neutron-deficient krypton isotopes are of particular interest for study due to their rapidly changing shapes with neutron number and the shape coexistence in the same nucleus, where the oblate and prolate shapes coexist within a very small energy range of a few hundreds keV.
The underlying reason is generally attributed to the abundance of low nucleon level densities,
or large ¡°shell gaps¡± for both prolate and oblate shapes at neutron/proton numbers 34, 36 and 38 in the Nilsson diagram.
Therefore, adding or removing only a few nucleons has a dramatic effect on the nuclear particle energies,
and consequently changes the ground state shape.

The experimental evidences associated with the shape coexistence for the neutron-deficient krypton isotopes
was first observed in the irregularity of the low-lying spectra of $^{74,76}$Kr more than three decades ago~\cite{Piercey1981,Piercey1982}.
More evidence on the prolate-oblate shape coexistence are found for $^{72,74}$Kr~\cite{Angelis1997,Chandler1997}.
Then the significantly reduced B(E2) value of the $2_1^+ \rightarrow 0_1^+$ transition for $^{72,74,76}$Kr were reported~\cite{Gorgen2005,Iwasaki2014}, indicating considerable shape mixing at lower spins.
More conclusive evidence of shape coexistence for even-even nuclei lies in
the identification of low-lying excited $0_2^+$ states which could be seen as the ¡°ground states¡± of other shape.
Two rotational bands may build on $0^+$ states with different deformations.
In 1999, an E0 transition at 508 keV is observed in $^{74}$Kr by means of combined conversion-electron and
$\gamma$-ray spectroscopy, confirming the existence of the expected low-lying isomeric $0_2^+$ state~\cite{Becker1999}.
This gives supports on the mixing between coexisting prolate and oblate shapes.
In 2003, the isometric $0_2^+$ state of $^{72}$Kr is identified~\cite{Bouchez2003}.
This state can be understood as the band head of a prolate rotational structure, which strongly supports the interpretation
that the ground state of $^{72}$Kr is oblate-deformed.
The systematics of excited $0^+$ states and the monopole transition strength in even-even nuclei $^{72-78}$Kr~\cite{Bouchez2003} were interpreted as evidence for an inversion of the ground-state deformation with decreasing neutron number: $^{78}$Kr and $^{76}$Kr are assumed to be prolate in their ground state, prolate and oblate configurations strongly mix in $^{74}$Kr, and an oblate shape becomes the ground state of $^{72}$Kr.
More direct evidence for the
coexistence of prolate and oblate shapes in $^{74,76}$Kr by means of Coulomb excitation is given where opposite signs are found for the quadrupole moments of the yrast and excited $2^+$ states in $^{74,76}$Kr~\cite{Clement2007}.
In 2015, $^{72}$Kr is studied with the total absorption spectroscopy technique, and its data can be interpreted
as a dominant oblate deformation or large oblate-prolate mixing in the ground state~\cite{Briz2015}.

Together with experimental efforts, various theories have been applied to elucidate what kinds of shape are involved and how they evolve,
including those employing Bohr¡¯s collective Hamiltonian \cite{Petrovici1983,Yao2013},
self-consistent triaxial mean-field models~\cite{Yamagami2001},
shell-model-based approaches ~\cite{Petrovici2000,Langanke2003},
beyond (relativistic) mean-field studies ~\cite{Bender2006,Yao2013,Rodrigues2014}, and
constrained Hartree-Fock-Bogoliubov (plus local Random-Phase-Approximation) calculations ~\cite{Girod2009,Sato2011},
the Total Routhian Surface method ~\cite{Bai2015},and  self-consistent Nilsson-like calculation~\cite{Zuker2015}.
In general, many of the global features of these Kr isotopes, such as coexistence of prolate and oblate shapes, their strong mixing at low
angular momentum, the deformation of collective bands, the low-spin spectra and the systematics of excitation energies and transition
strengths are reproduced.

The large shell gaps at prolate and oblate shapes at nucleon numbers 34, 36 and 38 cause the complicated shape evolution and shape coexistence in neutron-deficient Kr isotopes. If we excite the nucleus from another degree of freedom, e.g. the temperature, how will the shapes of these nuclei evolve with temperature? It is interesting to study how the shape evolves with temperature for neutron-deficient Kr isotopes from the nuclear structure point of view.
Additionally, $^{72}$Kr is one of the three major waiting points $^{64}$Ge, $^{68}$Se, and $^{72}$Kr
in the astrophysical rapid proton capture (rp) process, which powers type I X-ray bursts~\cite{Schatz2006}.
Since the environment of X-ray bursts has high temperature, it is also interesting to study the evolution of
$^{72}$Kr with temperature for astrophysical interest.

Usually, the shape deformations or superfluidity are expected to wash out in a heated nucleus~\cite{Egido2000}.
The equilibrated nucleus can be characterized by a temperature $T$ as an approximation to the microcanonical description.
This expectation can be understood in terms of the shell model
since by increasing temperature $T$ particles from levels below the Fermi surface are promoted to levels above it.
The basic thermal theory is developed by Ref. ~\cite{Bloch1958,Sauer1976}.
The shape transition at finite temperature is first studied in Ref. ~\cite{Lee1979}.
The finite temperature Hartree-Fock theory is developed ~\cite{Brack1974,Quentin1978} and
the dependence of nuclear shape transition on changes in the volume is studied by taking $^{24}$Mg as an example ~\cite{Yen1994}.
The finite temperature Hartree-Fock-Bogoliubov theory is formulated ~\cite{Goodman1981}
and then applied to the pairing and shape transitions in rare earth nuclei~\cite{Goodman1986}.
Using the finite range density dependent Gogny force and a large configuration space
within the framework of the finite-temperature Hartree-Fock-Bogoliubov theory~\cite{Egido2000,Egido2003},
various nuclei, including well quadrupole deformed nuclei, superdeformed nucleus, and octupole deformed nucleus,
gradually collapse to the spherical shape at certain critical temperatures ranging 1.3$\sim$2.7 MeV.
The temperature also affects the effective mass and the neutron skin~\cite{Yuksel2014}.

The covariant density functional theory (CDFT), which has achieved great success in describing ground-state
properties of both spherical and deformed nuclei all over the nuclear chart \cite{Ring1996,Vretenar2005,Meng2006,Meng2016}, is also applied to study the evolution of nuclear properties with temperature.
The finite-temperature relativistic Hartree-Bogoliubov theory~\cite{Niu2013} and relativistic Hartree-Fock-Bogoliubov theory~\cite{Long2015} for spherical nuclei are formulated, and used to study the pairing transitions in hot nuclei.
The relativistic Hartree-BCS theory is applied to study the temperature dependence of shapes and
pairing gaps for $^{166,170}$Er and rare-earth nuclei~\cite{Agrawal2000,Agrawal2001}.
A shape transition from prolate to spherical shapes is found at temperatures ranging 1.0$\sim$2.7 MeV.
Taking into account the unbound nucleon states,
the temperature dependence of the pairing gaps, nuclear deformation, radii, binding energies, entropy
are studied in the Dirac-Hartree-Bogoliubov (DHB) calculations~\cite{Bonche1984,Lisboa2016}.
It is also found the nuclear deformation disappears at temperatures $T = 2.0 - 4.0$ MeV.
When the temperature $T \geqslant 4$ MeV, the effects of the vapor phase that
take into account the unbound nucleon states become important.

It is clear that the sharp phase transitions obtained in the mean field
approach will be somewhat washed out when statistical fluctuations are considered.
The statistical fluctuations can be treated in
the spirit of the Landau theory ~\cite{Levit1984,Egido2003}, or
from a more fundamental point of view by using path integral techniques
like the static path approximation ~\cite{Alhassid1984,Rossignoli1994},
the shell model Monte Carlo ~\cite{Lang1993},
the particle number projected BCS ~\cite{Dang1993,Dang2007,Gambacurta2013}, or the
shell-model-like approach ~\cite{Liu2015}.

However, in the various previous studies, we didn't find any on the shape evolution of neutron-deficient Kr isotopes with temperature in the deformed relativistic framework.
So in our present work, we aim to investigate how the shape deformation changes when the temperature rises
for the shape coexistence region like $^{72,74}$Kr in the framework of CDFT.
For the shape coexistence phenomena, especially for the soft energy surfaces, the quantal fluctuations become important, therefore one needs the beyond-mean-field approach, such as the multiple-reference generator coordinate method (GCM)~\cite{Yao2013}, for quantitative descriptions.
However, as a first step towards this goal,
the self-consistent finite-temperature relativistic mean field with BCS approach for axially deformed nuclei based on the point-coupling density functional is developed in our paper for the first time, and it is used to investigate the
free energy curves, the quadroupole deformations, the pairing correlations
as functions of the temperatures for isotopes $^{72,74}$Kr.
The evolution of the shapes and single-particle spectrum will be discussed.
Considering the effects of the vapor phase become important when $T\geqslant 4.0$ MeV in the DHB calculations~\cite{Lisboa2016},
we limit the temperature range to 0-4 MeV in our study.

\section{Theoretical framework}
\label{sec2}

The starting point of the CDFT is an effective Lagrangian density
with zero-range point-coupling interaction between nucleons:
\begin{eqnarray}\label{Eq:Lagrangian}
 \mathcal{L}&=& \bar\psi(i\gamma_\mu\partial^\mu-m)\psi \nonumber\\
            && -\frac{1}{2}\alpha_S(\bar\psi\psi)(\bar\psi\psi)
               -\frac{1}{2}\alpha_{V}(\bar\psi\gamma_\mu\psi)(\bar\psi\gamma^\mu\psi)
               -\frac{1}{2}\alpha_{TV}(\bar\psi\vec\tau\gamma_\mu\psi)\cdot(\bar\psi\vec\tau\gamma^\mu\psi) \nonumber\\
            && -\frac{1}{3}\beta_S(\bar\psi\psi)^3-\frac{1}{4}\gamma_S(\bar\psi\psi)^4
               -\frac{1}{4}\gamma_V[(\bar\psi\gamma_\mu\psi)(\bar\psi\gamma^\mu\psi)]^2 \nonumber\\
            && -\frac{1}{2}\delta_S\partial_\nu(\bar\psi\psi)\partial^\nu(\bar\psi\psi)
               -\frac{1}{2}\delta_V\partial_\nu(\bar\psi\gamma_\mu\psi)\partial^\nu(\bar\psi\gamma^\mu\psi) \nonumber\\
            && -\frac{1}{2}\delta_{TV}\partial_\nu(\bar\psi\vec\tau\gamma_\mu\psi)\cdot\partial^\nu(\bar\psi\vec\tau\gamma^\mu\psi)\nonumber\\
            && -\frac{1}{4}F^{\mu\nu}F_{\mu\nu}  - e\bar\psi\gamma^\mu\frac{1-\tau_3}{2}\psi A_\mu,
\end{eqnarray}
which includes the free nucleons term, the four-fermion point-coupling terms,
the higher-order terms which are responsible for the effects of medium dependence,
the gradient terms which are included to simulate the effects of finite range,
and the electromagnetic interaction terms.
The isovector-scalar channel is neglected.
The Dirac spinor field of the nucleon is denoted by $\psi$, and the nucleon mass is $m$.
$\vec\tau$ is the isospin Pauli matrix, and $\Gamma$ generally denotes the 4$\times$4 Dirac matrices
including $\gamma_\mu$, $\sigma_{\mu\nu}$ while
Greek indices $\mu$ and $\nu$ run over the Minkowski indices 0, 1, 2, and 3.
$\alpha$,$\beta$,$\gamma$,and $\delta$ with subscripts $S$ (scalar),$V$ (vector),$TV$ (isovector) are coupling constants (adjustable parameters) in which $\alpha$ refers to the four-fermion term, $\beta$ and $\gamma$ respectively to the third- and fourth-order terms,
$\delta$ the derivative couplings.

Following the prescription in Ref.~\cite{Goodman1981} where the BCS limit of finite-temperature
Hartree-Fock Bogoliubov equations is derived, we obtain the finite-temperature CDFT + BCS equation.
The finite-temperature Dirac equation for single nucleons reads
\begin{equation}\label{Eq:Dirac-PC}
  [\gamma_\mu(i\partial^\mu-V^\mu)-(m+S)]\psi_k=0,
\end{equation}
where $m$ is the nucleon mass.  $\psi_k(\bm{r})$ denotes the Dirac spinor field of a nucleon.
The scalar $S(\bm{r})$ and vector potential $V^\mu(\bm{r})$ are
\begin{equation}\label{Eq:S}
S(\bm{r})  =\alpha_S\rho_S+\beta_S\rho^2_S+\gamma_S\rho^3_S+\delta_S\triangle\rho_S,
\end{equation}
\begin{eqnarray}\label{Eq:V}
V^\mu (\bm{r}) &=&\alpha_Vj^\mu_V +\gamma_V (j^\mu_V)^3 +\delta_V\triangle j^\mu_V\nonumber\\
               & &+\tau_3\alpha_{TV} \vec{j}^\mu_{TV}+ \tau_3\delta_{TV}\triangle \vec{j}^\mu_{TV}+ e A^\mu
\end{eqnarray}
respectively.
The isoscalar density, isoscalar current and isovector current are denoted by $\rho_S$, $j^\mu_V$, and $\vec{j}^\mu_{TV}$ respectively, and have the following form,

\begin{eqnarray}
\rho_S (\bm{r})      &=& \sum \limits_{k} \bar\psi_k(\bm{r}) \psi_k(\bm{r}) [v_k^2 (1-2 f_k)+f_k], \\
j^\mu_V (\bm{r})     &=& \sum \limits_{k} \bar\psi_k(\bm{r}) \gamma^\mu \psi_k(\bm{r}) [v_k^2 (1-2 f_k)+f_k], \\
\vec{j}^\mu_{TV} (\bm{r}) &=& \sum \limits_{k} \bar\psi_k(\bm{r}) \vec{\tau} \gamma^\mu \psi_k(\bm{r}) [v_k^2 (1-2 f_k)+f_k].
\end{eqnarray}
$f_k$ is
the thermal occupation probability of quasiparticle states, which has the form $f_k=1/(1+e^{\beta E_k})$.
$E_k$ is the quasiparticle energy for single particle (s.p.) state $k$,
$E_k = [(\epsilon_k-\lambda)^2 +( \Delta_k)^2]^{\frac{1}{2}}$.
$\beta = 1/(k_BT)$ where $k_B$ is the Boltzmann constant.
The BCS occupation probabilities $v_k^2$ and related $u_k^2=1-v_k^2$ are obtained by
\begin{eqnarray}\label{eq:occ}
v_k^2   &=&\frac{1}{2} (1- \frac{\epsilon_k-\lambda}{E_k}) \\
u_k^2   &=&\frac{1}{2} (1+ \frac{\epsilon_k-\lambda}{E_k}).
\end{eqnarray}
$\Delta_k$ is the pairing gap parameter, which satisfies the gap equation at finite temperature.
\begin{equation}
  \Delta_k = - \frac{1}{2} \sum_{k'>0} V^{pp}_{k\bar{k} k' \bar{k}'} \frac{\Delta_{k'}}{ E_{k'}} (1-2f_{k'}).
\end{equation}
The particle number $N_q$ is kept by $N_q= 2 \sum \limits_{k>0} [v_k^2 (1-2 f_k)+f_k]$.

Here we take the $\delta$ pairing force $V(\bm{r})=V_q\delta(\bm{r})$, where $V_q$ is the pairing strength parameter for neutrons or protons.
A smooth energy-dependent cutoff weight $g_k$ is introduced to simulate the effect of finite range
and determined by an approximate condition $\sum \limits_{k} 2 g_k= N_q + 1.65 N^{2/3}_q $ related to the particle number $N_q$.

The internal binding energy for the nuclear system $E$ is
\begin{equation}
  E = E_{\rm part} + E_{\rm int} + E_{\rm pair} + E_{\rm c.m.} -A M,
\end{equation}
where $E_{\rm part}$ is the total single-particle energy,
\begin{eqnarray}\label{eq:ed}
E_{\rm part} &=& 2 \sum \limits_{k} \epsilon_k [v_k^2 (1-2 f_k)+f_k];
\end{eqnarray}
$E_{\rm int}$ is the mean-field potential energy,
\begin{eqnarray}
E_{\rm int}  &=&
-\int d\bm{r} \left[ \frac{\alpha_S}{2}\rho_S^2+\frac{\beta_S}{3}\rho_S^3+\frac{\gamma_S}{4}\rho_S^4+\frac{\delta_S}{2}\rho_S\triangle \rho_S \right.\nonumber \\
  &&+ \frac{\alpha_V}{2}j_\mu j^\mu + \frac{\gamma_V}{4}(j_\mu j^\mu)^2 + \frac{\delta_V}{2}j_\mu\triangle j^\mu  \nonumber \\
  &&\left.+ \frac{\alpha_{TV}}{2}\vec{j}^\mu_{TV}(\vec{j}_{TV})_\mu
    + \frac{\delta_{TV}}{2}\vec{j}^\mu_{TV} \triangle (\vec{j}_{TV})_{\mu}\right] \nonumber \\
  && -\int d\bm{r} \left[\frac{1}{4}F_{\mu\nu}F^{\mu\nu}-F^{0\mu}\partial_0 A_\mu+\textit{e}A_\mu j_p^\mu \right] ;
\end{eqnarray}
$E_{\rm pair}$ is the pairing energy,
\begin{equation}\label{eq:epair}
    E_{\rm pair}= - \sum \limits_{k} \Delta_k u_k v_k (1-2 f_k);
\end{equation}
and $E_{\rm c.m.}$ is the central of mass correction energy.

The internal binding energy $E$ at different quadroupole deformation $\beta_2$ can be obtained by applying constraints.
The entropy of the system is evaluated by
\begin{equation}\label{eq:S}
S=-k_B \sum \limits_{k} [f_k{\rm ln}f_k +(1-f_k){\rm ln}(1-f_k)],
\end{equation}
and the free energy is $F=E-TS$.
For convenience, the temperature is used as $k_BT$ in unit of MeV and
the entropy is used as $S/k_B$ unitless.
The specific heat is defined as the derivative of the excitation energy by
\begin{equation}
C_{\rm v}=\partial E^*/\partial T
\end{equation}
where $E^*(T)=E(T)-E(T=0)$ is the internal excitation energy, and
$E(T)$ is the internal binding energy for the global minimum state in the free energy curve at certain temperature $T$.

\section{Results and discussion}

The point-coupling density functional parameter set PC-PK1 is used in our calculation
due to its success in the description of finite nuclei for both ground state and low-lying excited states~\cite{Zhao2010}.
The pairing correlations are taken into account by the $\delta$ force BCS method with a smooth cutoff factor.
The value of the pairing strength for neutrons (protons) $V_q$ is taken from Ref.~\cite{Zhao2010},
that is, -349.5 (-330.0) MeV fm$^3$.
A set of axial harmonic oscillator basis functions with 20 major shells is used.

\begin{center}
\includegraphics[scale=0.5]{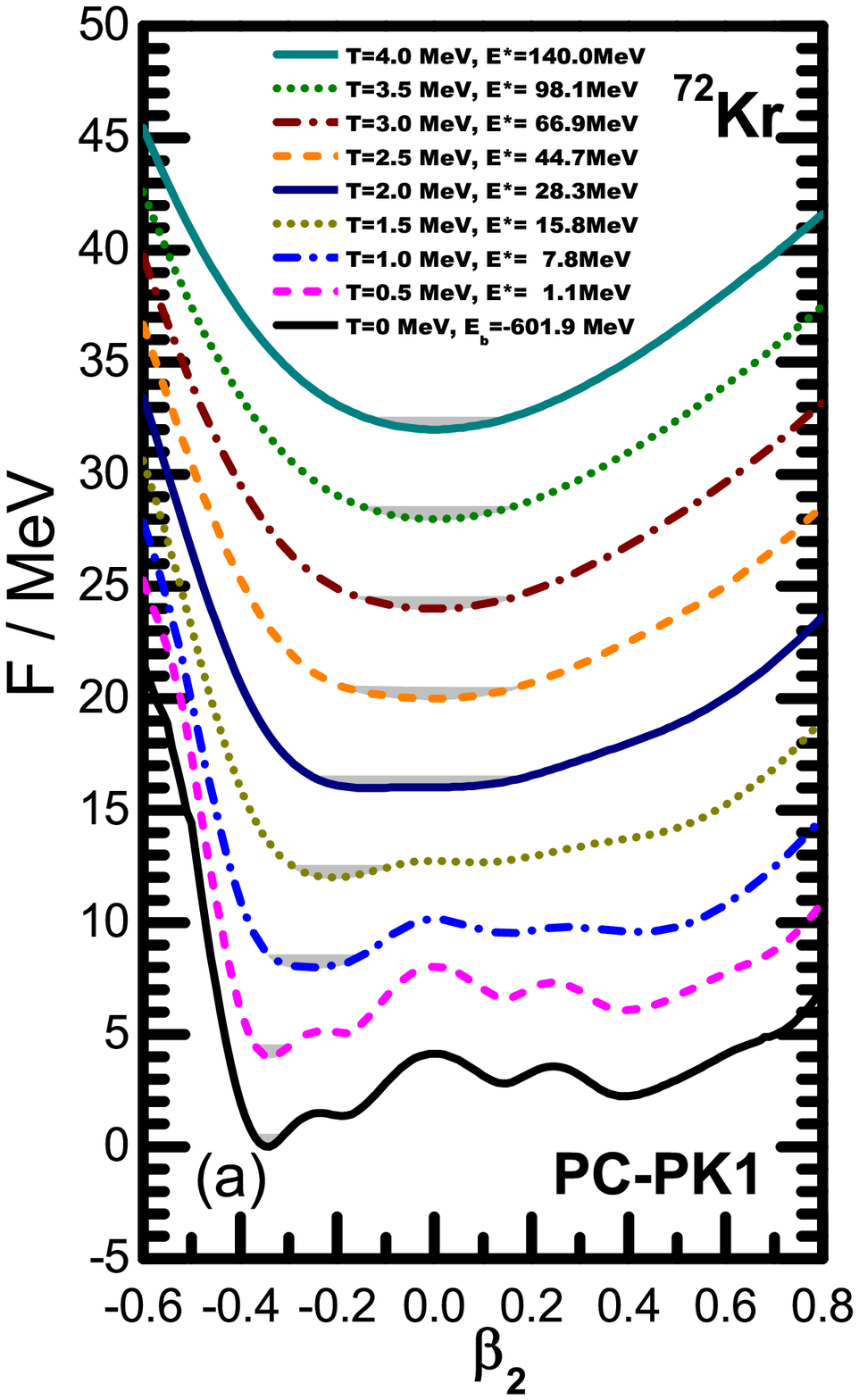}
\includegraphics[scale=0.5]{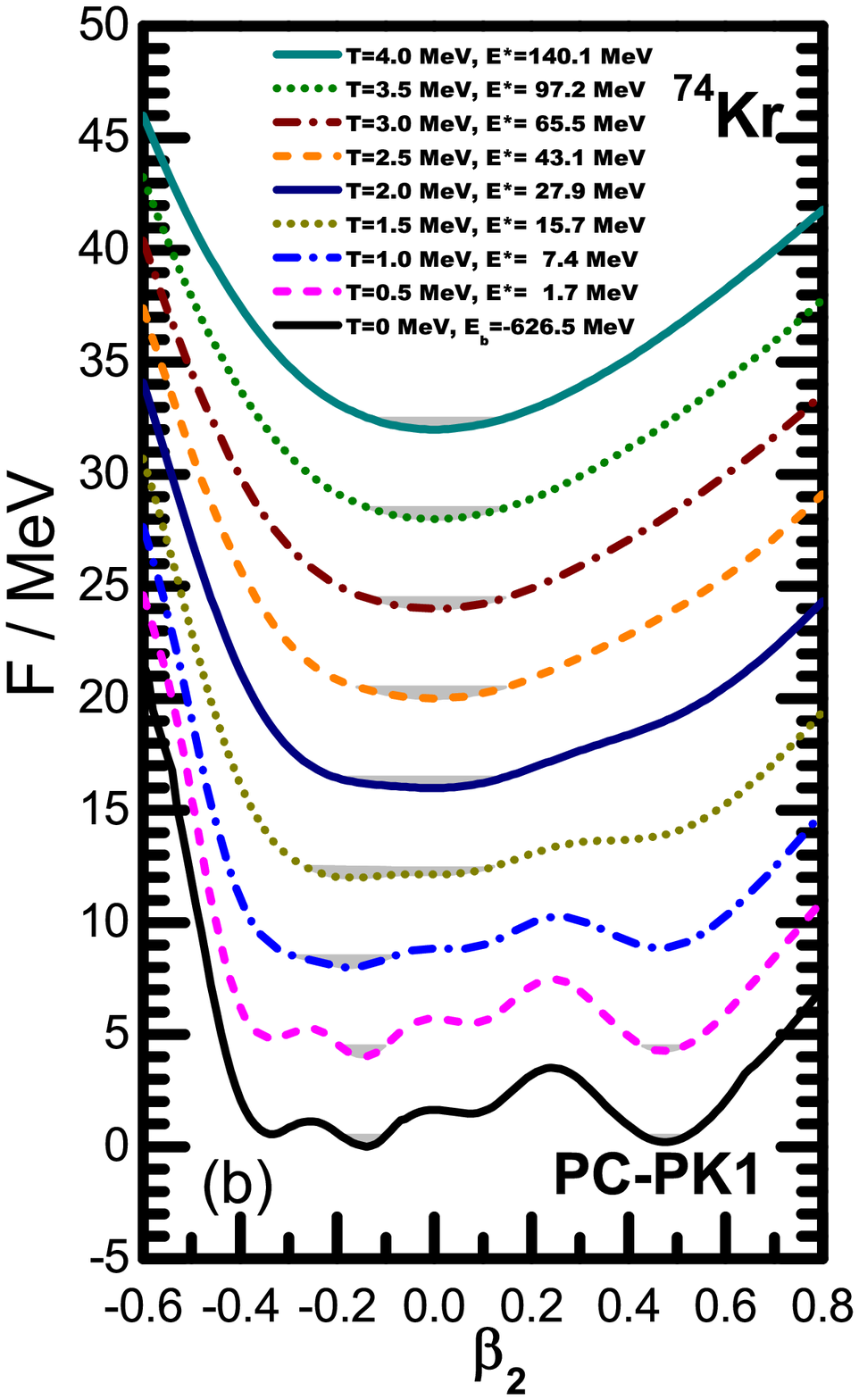}\\
\figcaption{\label{pes}
The relative free energy curves for $^{72}$Kr (a) and $^{74}$Kr (b)
at different temperatures from 0 to 4 MeV with the step 0.5 MeV
obtained by the constrained CDFT+BCS calculations using PC-PK1 energy density functional.
The ground state free energy at zero temperature is set as zero, and it is
shifted up by 4 MeV for every 0.5 MeV temperature rise. The absolute ground state binding energy as well as the excitation energies at higher temperatures are shown in labels. The shaded area are marked for states whose energies
are no more than 0.5 MeV above the global minimum of the free energy curves at the corresponding temperature.
}
\end{center}


The relative free energies as functions of $\beta_2$ at different temperatures from 0 to 4 MeV for isotopes $^{72,74}$Kr are plotted in Fig.~\ref{pes} side by side.
In order to see the energy curves clearly, the free energy of the ground state is chosen as zero, and it is
shifted up by 4 MeV for every 0.5 MeV temperature rise.
Let's first analyze the behavior of the free energy curves for $^{72,74}$Kr at zero temperature.
For $^{72}$Kr, there are four local minima at $\beta_2$= -0.34, -0.19, 0.14, and 0.39.
The corresponding energies relative to the ground state at $\beta_2$=-0.34 read 0, 1.36, 2.82, and 2.24 MeV, respectively.
So at zero temperature, the energy curve is not flat, and the global minimum is well distinguished from other local minima.
For triaxial calculations with the same parameter set~\cite{Yao2013},
the global minimum locates at $|\beta_2|$=0.35, $\gamma = 60^\circ$
(equivalent to $\beta=-0.35$ in axial deformation system),
while the minimum at $|\beta_2|$=0.19, $\gamma = 60^\circ$ is actually a saddle point in ($\beta_2$, $\gamma$) plane.
The quadrupole deformation of both calculations are in consistent with the experimental value $|\beta_2|$=0.330~\cite{Gade2005}.

For $^{74}$Kr, similar as $^{72}$Kr, also four local minima at $\beta_2$= -0.36, -0.14, 0.07, and 0.47 exist.
Although there is a 3.3 MeV barrier, well separating the prolate minimum at $\beta_2=0.47$ and other minima,
the two of them with lower energies are located at $\beta_2$= -0.14 and 0.47 with only an energy difference of 0.20 MeV,
which shows the possible shape coexistence in this nucleus.
In the present calculation, state at $\beta_2$ =-0.14 has the lowest energy.
We may notice that this global minimum differs from  that of the triaxial calculation~\cite{Yao2013}, which prefers the other minimum at $|\beta_2|$ = 0.50, $\gamma = 0^\circ$.
Since different kinds of pairing forces are applied, delta pairing force in our case and separable pairing force
in Ref.~\cite{Yao2013}, it is not surprising that the global minimum changes with such a small energy difference.
In such a case where the energies of prolate and oblate shape are so close to each other,
a proper treatment like the GCM theory is in need for describing the mixing between prolate and oblate shapes.
However, from our mean-field calculation, the potential curve qualitatively supports the physical picture of
such shape coexistence, which is consistent with the experimental data~\cite{Bouchez2003,Clement2007}.
The experimental deformation for $^{74}$Kr ground state reads $|\beta_2|=$0.419~\cite{Raman2001}.
The opposite signs of spectroscopic quadrupole moments are found for the ground-state bands and
the bands based on excited $0_2^+$ states, with 508 keV energy difference between the corresponding bandheads~\cite{Clement2007}.
The assumption of maximum mixing between a strongly prolate ($\beta_2 \sim$ 0.4) and
a weaker oblate configuration ($\beta_2 \sim$ -0.1) for the $0^+$ states of $^{74}$Kr
is supported by the two-level mixing model, which is in consistency with the experimental data~\cite{Clement2007}.
For non-relativistic Total Routhian Surface calculations~\cite{Bai2015},
the ground state deformations for $^{72}$Kr and $^{74}$Kr are -0.333 and 0.381 ($\gamma=2^\circ$) respectively.

With the temperature increasing, the barrier that separates the prolate and oblate shapes becomes weaker
and finally vanishes at $T \sim$ 2.1 MeV and 1.7 MeV for $^{72}$Kr and $^{74}$Kr respectively, where large flat curve segments are developed, indicated by the shaded area.
For example, the states with energy no more than 0.5 MeV higher than the minimum at $T$=1.8 MeV for $^{72}$Kr are located in
the deformation interval $-0.27 \leqslant \beta_2 \leqslant 0.17$.
When the temperature rises,
more nucleons are distributed to single-particle levels with high energies,
which smears the energy differences at different deformations, and thus a soft area is developed.
It is clear that the nuclei $^{72,74}$Kr share similar soft free energy areas at high temperatures,
since the difference between the two nuclei are also smeared by the high temperature.

Before the free energy potential curve for $^{72}$Kr and $^{74}$Kr becomes soft,
the two nuclei start to evolve with temperature from different ground state properties.
For $^{72}$Kr, the nucleus changes from oblate to a
soft curve with the spherical shape as its minimum at $T\sim 2.1$ MeV.
Instead of a well localized minimum at zero temperature in $^{72}$Kr,
$^{74}$Kr has two shapes, one oblate and the other prolate, with similar energies for low temperatures.
With the temperature increasing, the energy difference between the oblate and prolate minima becomes larger gradually,
e.g. the prolate minimum at $\beta_2 \sim 0.47$ is 0.5 MeV higher than the oblate minimum at $\beta_2 \sim -0.14$
at $T=0.75$ MeV, and 1 MeV higher at $T=1.05$ MeV.
At the same time, the local minimum at $\beta_2 \sim 0.47$ becomes shallower relative to its neighbor states
and eventually vanishes at $T=1.45$ MeV.
With the temperature further increasing, the nucleus $^{74}$Kr changes from oblate to a soft curve with the spherical global minimum at $T=1.7$ MeV as $^{72}$Kr.

\begin{center}
\includegraphics[scale=0.6]{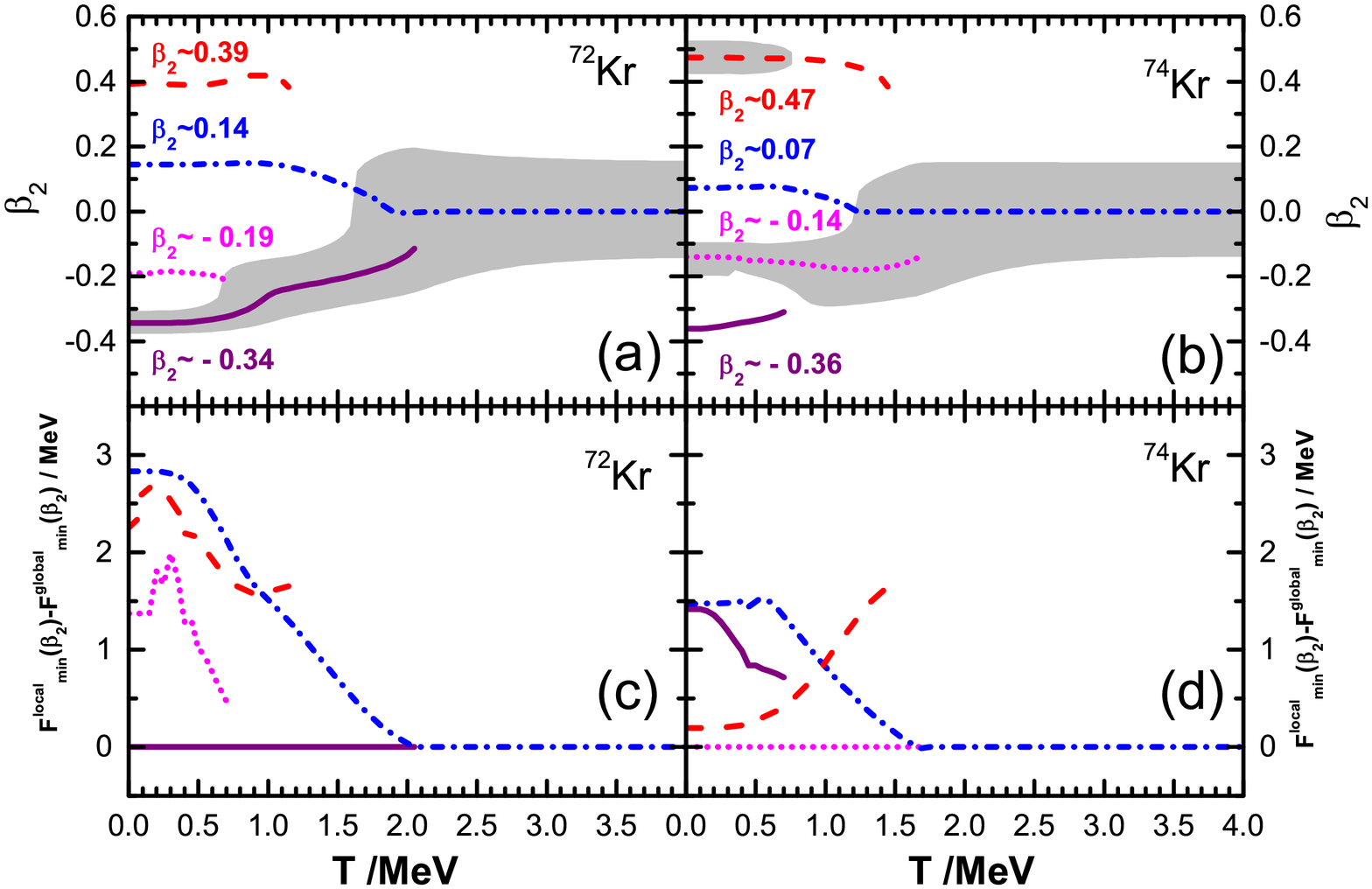}
\figcaption{\label{basic}
The local minima deformation $\beta_2$ and their free energies relative to the global minimum (in MeV)
as functions of temperature (in MeV) for $^{72}$Kr (a,c) and $^{74}$Kr(b,d), obtained by the constrained CDFT+BCS calculations using PC-PK1 energy density functional.
}
\end{center}

To analyze the shape properties of the nuclei as functions of temperatures in more detail,
the evolutions of deformation and relative energy to the global minimum of all minima with temperatures for $^{72,74}$Kr
are plotted in Fig.~\ref{basic}.
Both $^{72}$Kr and $^{74}$Kr have four local minima, and the evolutions of these four local minima with temperature are similar for $^{72}$Kr and $^{74}$Kr. However, the relative energies between different minima are different for these two nuclei, which leads to the different evolution behavior of the nuclear shape.
For $^{72}$Kr, the energies of other minima are much higher than that of the global minimum,
normally above 1.3 MeV, which can be seen in Fig.~\ref{basic}(c).
This situation isolates the global minimum at $\beta_2 \sim -0.34$.
This oblate minimum gradually evolves to spherical with two quick deformation changes,
one at $T \sim 0.9$ MeV, the other at $T \sim 2.1$ MeV.
For $^{74}$Kr, the oblate and prolate minima at $\beta_2\sim -0.14$ and $\beta_2 \sim 0.47$
compete strongly for low temperatures.
It should be noted that the local minimum $\beta_2 \sim -0.14$ becomes the global minimum,
not the more oblate minimum $\beta_2 \sim -0.36$, which is corresponding to the global minimum of $^{72}$Kr.
This global minimum stably locates at $\beta_2=-0.14$ for $T\leqslant 0.5$ MeV,
and slightly wobbles around $\beta_2=-0.14$ for higher temperatures $0.5 \leqslant T \leqslant 1.7$ MeV.
In Fig.~\ref{basic}(d), the relative free energy differences between the local minima and
global minimum for $^{74}$Kr share similar behavior as those of $^{72}$Kr shown in Fig.~\ref{basic}(c),
but with small amplitudes.
It can be seen in Fig.~\ref{basic} that
$^{72}$Kr experiences one continuous deformation change at $T=0.9$ MeV and one abrupt deformation change at $T=2.1$ MeV while
$^{74}$Kr experiences one abrupt shape change at $T=1.7$ MeV.

\begin{center}
\includegraphics[scale=0.6]{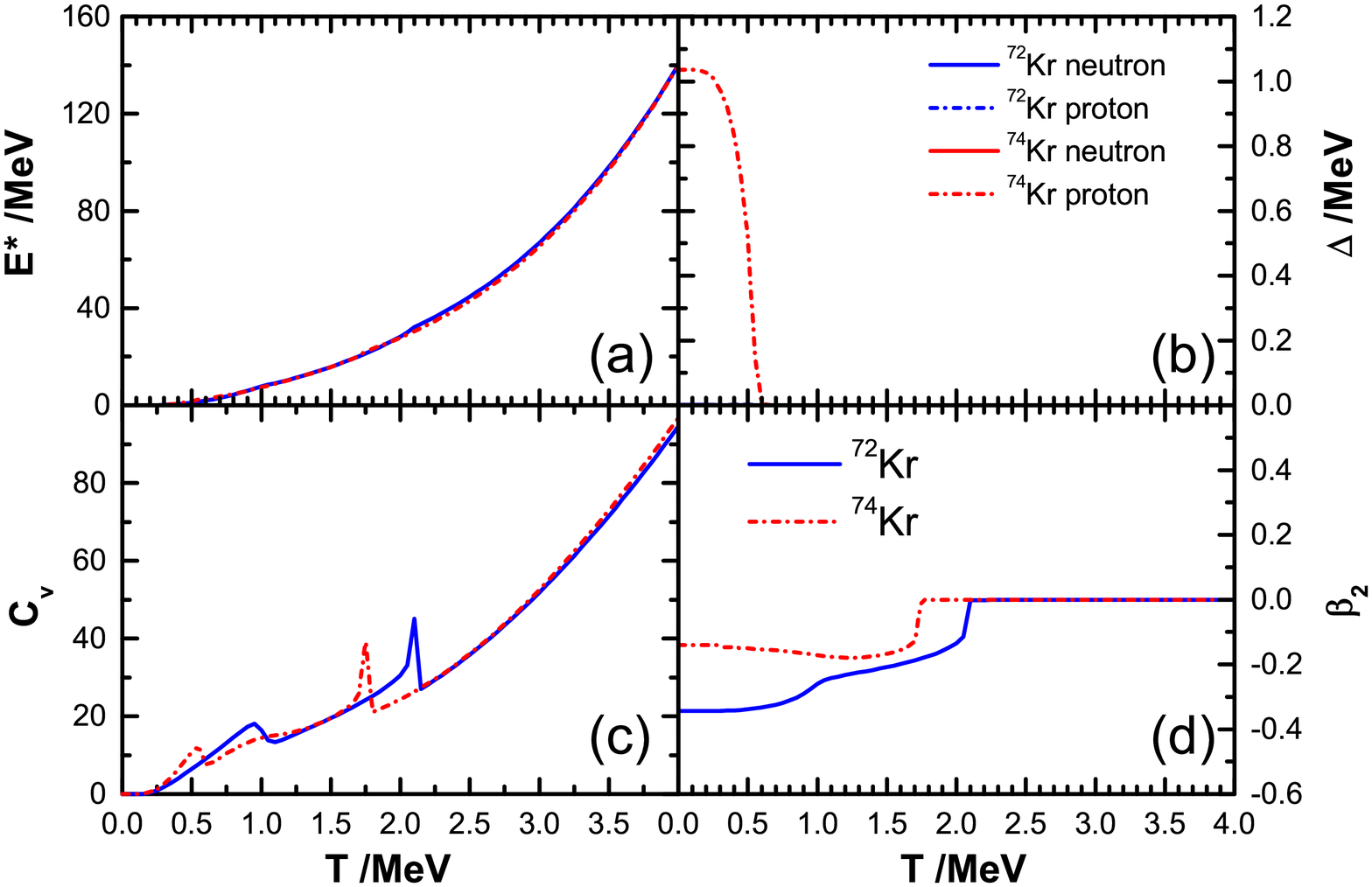}
\figcaption{\label{ExCv}
The excitation energy $E^*$ (in MeV) (a), pairing gaps $\Delta_n$, $\Delta_p$ (in MeV) (b), the specific heat $C_{\rm v}$ (c),
and the global minimum deformation $\beta_2$ (d) as functions of temperature (in MeV) for $^{72,74}$Kr, obtained by the constrained CDFT+BCS calculations using PC-PK1 energy density functional.}
\end{center}

Additionally, the bulk properties of the nuclei $^{72,74}$Kr as functions of temperatures, including
the excitation energy, the pairing gaps, the specific heat as well as the global minimum deformation
are shown in Fig.~\ref{ExCv}.
In Fig.~\ref{ExCv}(a), the relative excitation energies $E^*$ for $^{72,74}$Kr are
very similar while its values are shown in Fig.~\ref{pes} labels.
For pairing gaps, it is well-known that,
at a fixed temperature, the pairing gaps may vary a lot at different deformations, depending on the
specific neutron or proton single-particle level structures.
In Fig.~\ref{ExCv}(b), both the neutron and proton pairing gaps near $\beta_2 \sim -0.34$ for $^{72}$Kr
as well as the neutron pairing gap near $\beta_2 \sim -0.14$ for $^{74}$Kr are zero due to the large shell gaps
(cf. Fig.~\ref{Kr72sp} and ~\ref{Kr74sp}).
The proton pairing gap near $\beta_2 \sim -0.14$ for $^{74}$Kr gradually decrease to nearly zero,
basically following the rule $T_c = 0.6\Delta_p(0)$, where $T_c=0.60$ MeV is the critical temperature for a pairing transition
and $\Delta_p(0)=1.03$MeV is the proton pairing gap at zero temperature,
since the deformation and associated single particle levels for this minimum change little with rising temperature.
In Fig.~\ref{ExCv}(c), two discontinuities for $^{72,74}$Kr can be found respectively.
For $^{72}$Kr, the discontinuities at $T= 0.9$ MeV and $2.1$ MeV in Fig.~\ref{ExCv}(c) match
two deformation changes in Fig.~\ref{ExCv}(d).
For $^{74}$Kr, the discontinuity at $T= 0.6$ MeV matches the proton pairing transition temperature in Fig.~\ref{ExCv}(b)
while the discontinuity at $T= 1.7$ MeV matches the deformation change in Fig.~\ref{ExCv}(d).
Based on these figures,
the specific heat is a good signature in the search for pairing transition or shape change.
However, in experiment the specific heat usually exhibits a more smooth behavior
as compared to the sharp discontinuity obtained here.
This is attributed to the finite size of the nucleus and, therefore,
realistic description of statistical and quantal fluctuations.

\begin{center}
\includegraphics[scale=0.3]{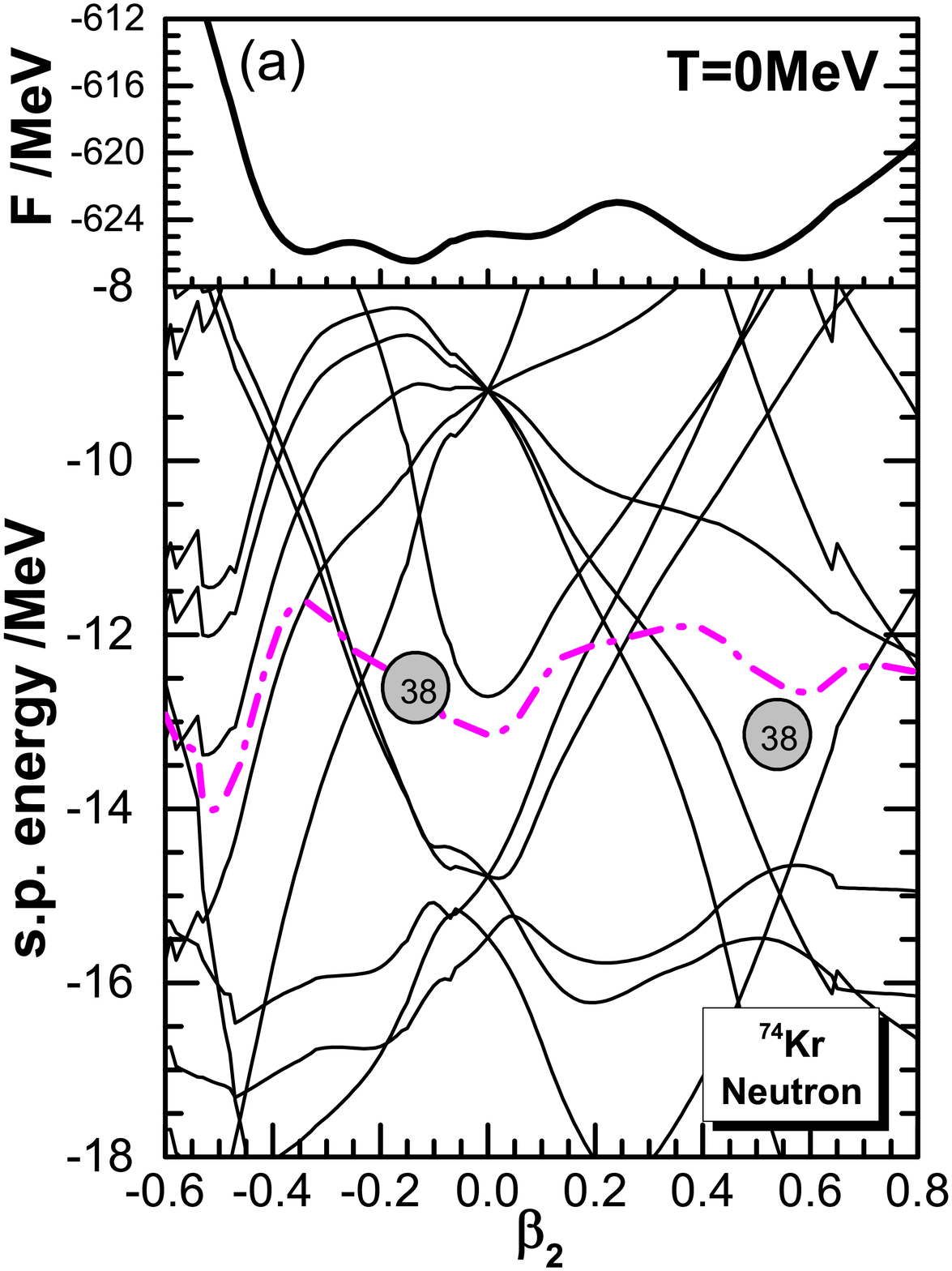}%
\includegraphics[scale=0.3]{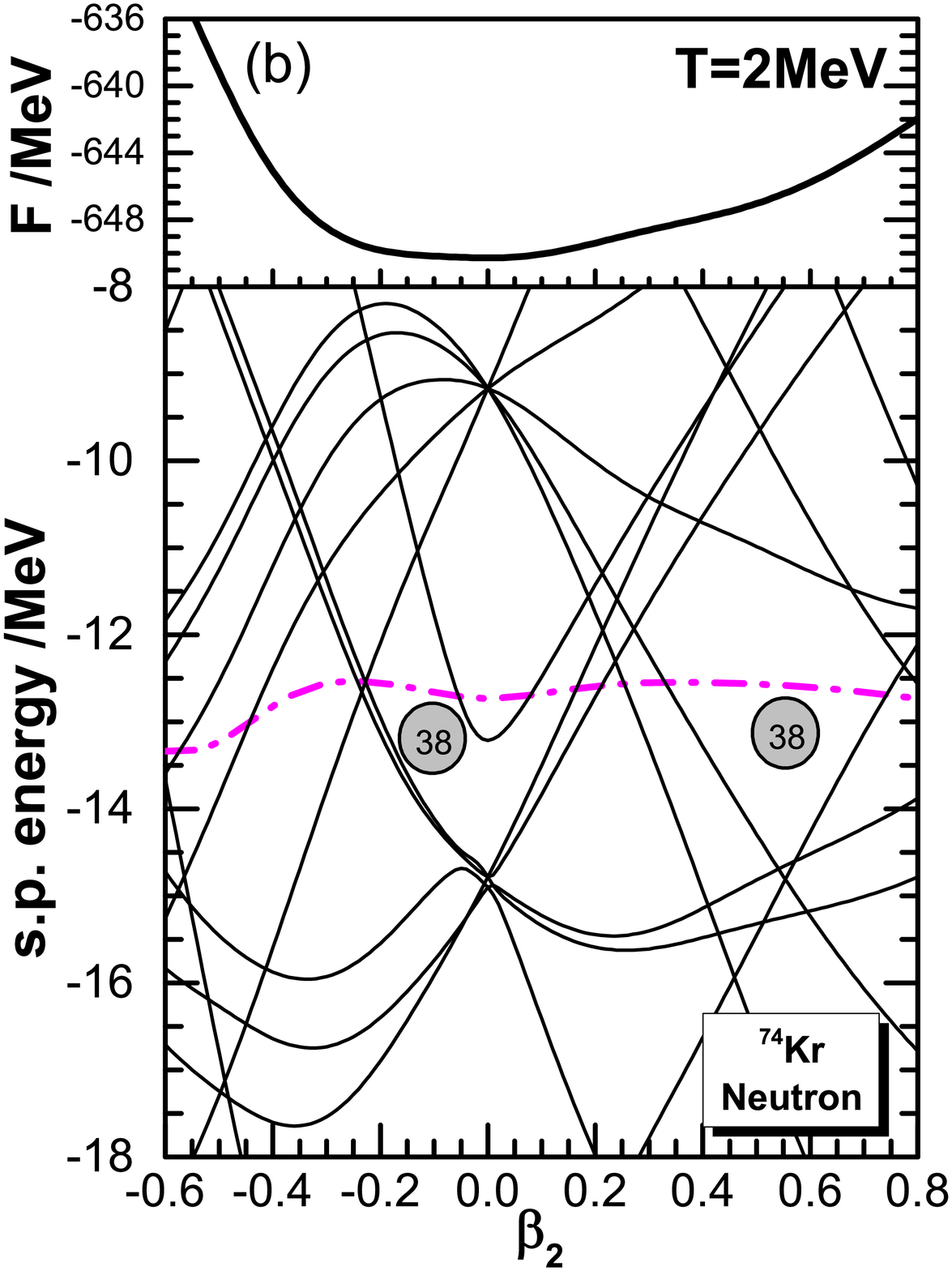}%
\figcaption{\label{Kr74T}
Neutron single-particle levels and energy potential curves as a function of deformation $\beta_2$ for the nucleus $^{74}$Kr at $T$=0 (a), and 2 MeV (b),  obtained by the constrained CDFT+BCS calculations using PC-PK1 energy density functional.
The dash-dot lines denote the corresponding Fermi surfaces.
}
\end{center}

For a more microscopic study, we check the temperature effects on the shell structure, so
the Nilsson diagrams for neutrons of $^{74}$Kr at temperatures $T=$0, and 2 MeV are plotted in Fig.~\ref{Kr74T},
together with the free energy curves at corresponding temperatures.
It can be seen that the Nilsson diagrams are almost the same at different temperatures, which shows that the temperature has small effect on the single particle energy at the same deformation.
The Fermi energy is largely modified since the occupation of s.p. levels changes a lot by increasing temperature.
Such effect is similar to shape transition with increasing nucleon numbers.
For shape transition between nuclei, due to the abundance of low nucleon level densities, or subshell gaps in the Nilsson diagram,
adding or removing only a few nucleons might have a dramatic effect on
the particle energies and consequently change the ground state shape.
Here the temperature promotes nucleons from levels below the Fermi surface to levels above it,
crossing the pronounced subshell gaps at nucleon numbers 38 (oblate $\beta_2 \sim$ -0.14 or prolate $\beta_2 \sim$ 0.47)
in Fig.~\ref{Kr74T},
and may demonstrate the changes of dominant shapes in one nucleus.
Through the alignment between the energy potential curve and the s.p. structure,
it is very clearly seen that the gaps that the Fermi surface goes through
coincide with the local minima in the potential curves at zero temperature.
The existence of these intruder states which form the gap structure in the Nilsson diagram is responsible for the deformed ground state.
However, at $T=2$ MeV, the shell structure of the Nilsson diagram doesn't influence the position of the minimum any more due to the diffusion of nucleons on the s.p. levels, and the spherical shape is always preferred. With temperature increasing, the shell effects on the nucleus gradually fades away.

\begin{center}
\includegraphics[scale=0.65]{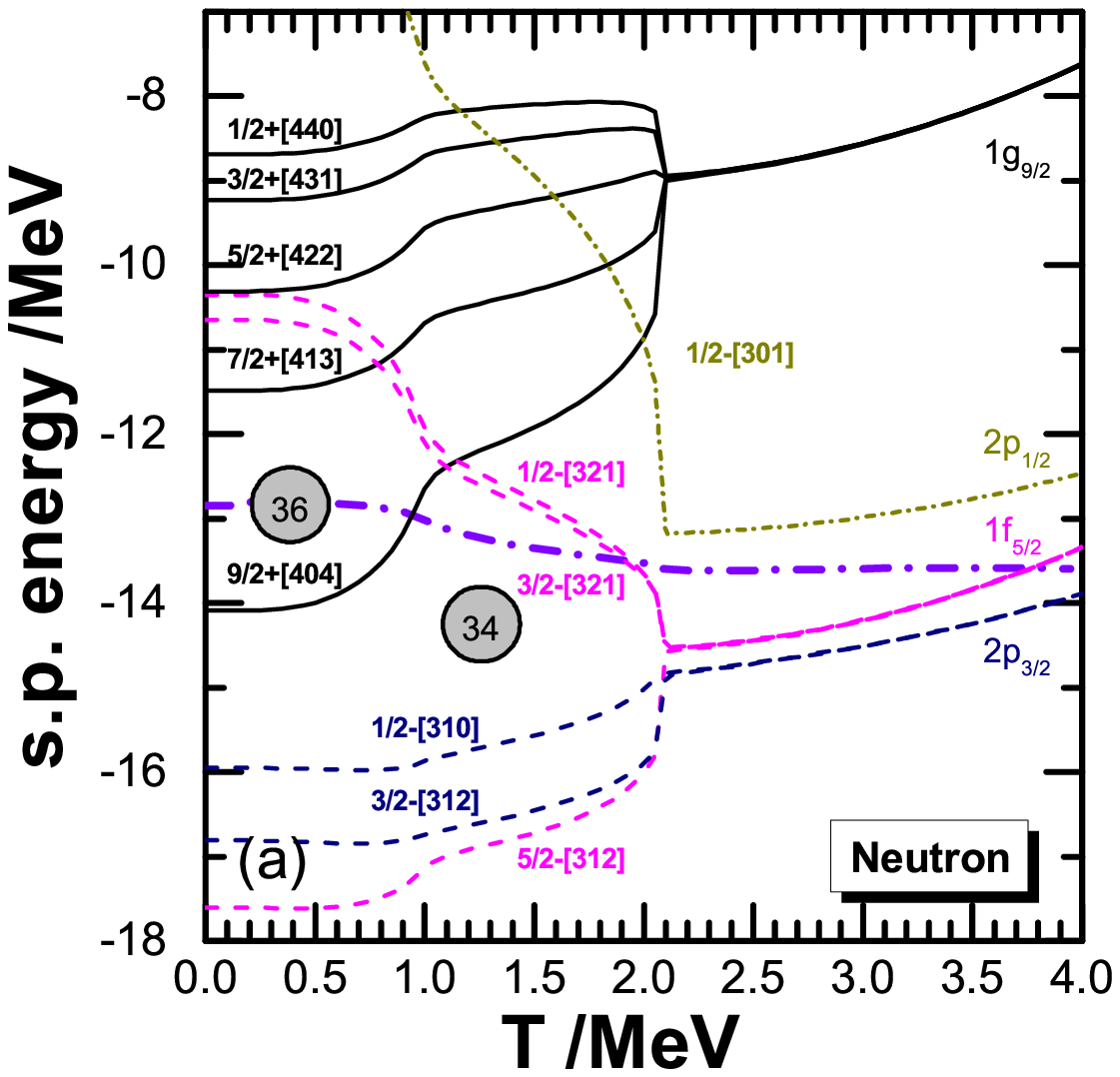}%
\includegraphics[scale=0.65]{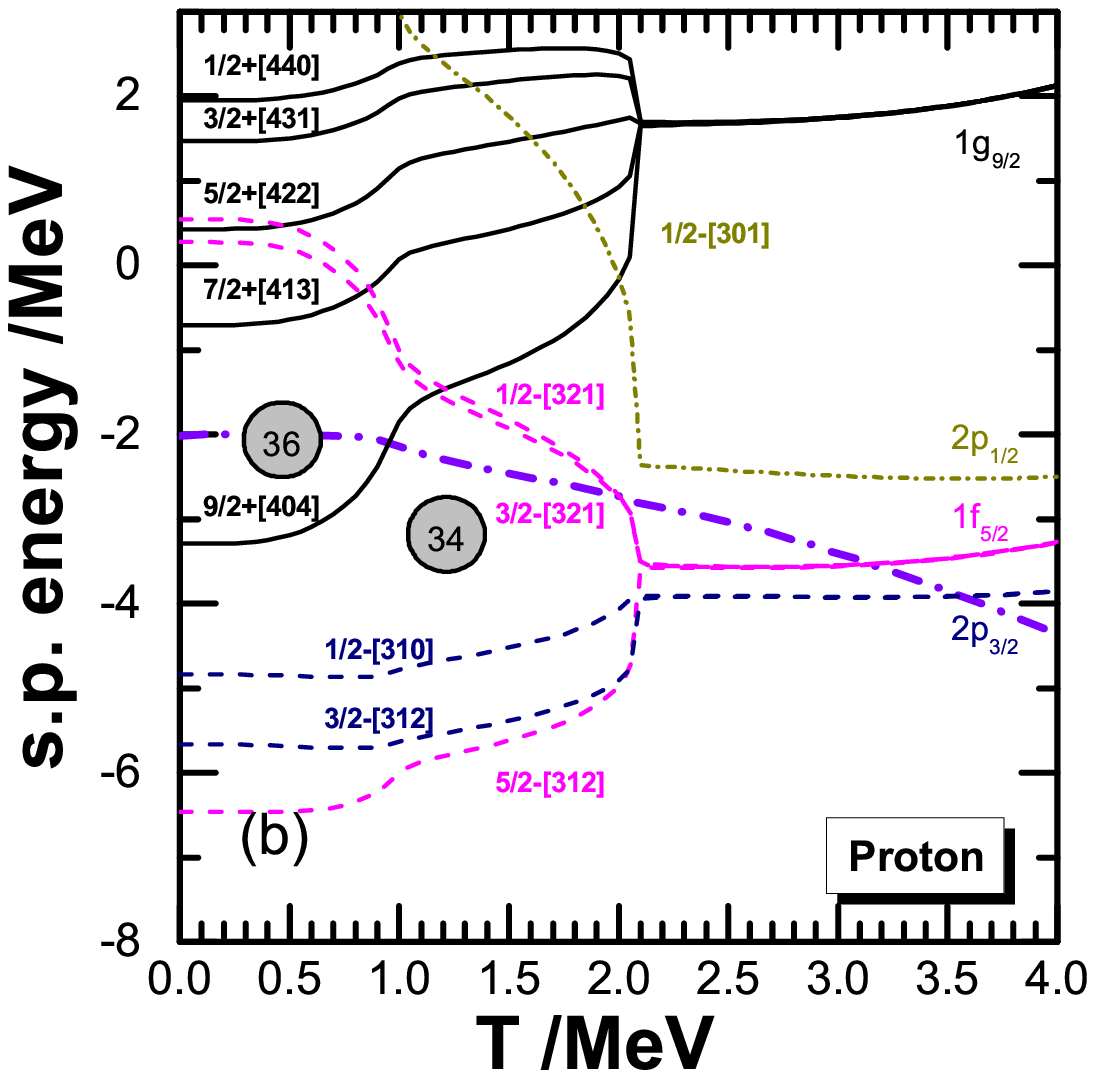}\\
\figcaption{\label{Kr72sp}
Neutron (a) and proton (b) single-particle levels as a function of temperature (in MeV) for the nucleus $^{72}$Kr,  obtained by the constrained CDFT+BCS calculations using PC-PK1 energy density functional.
The dash-dot lines denote the corresponding Fermi surfaces.
The levels near the Fermi surface are labeled by Nilsson notations $\Omega$$\pi$[$N$$n_z$$m_l$]
of the first leading component at zero temperature.
}
\end{center}


In the following Fig.~\ref{Kr72sp} and Fig.~\ref{Kr74sp}, we plot the the s.p. levels of neutrons and protons at the global minimum obtained from our mean-field calculation as a function of temperature for $^{72}$Kr and $^{74}$Kr respectively.
For $^{72}$Kr in Fig. ~\ref{Kr72sp},
the s.p. level evolutions for neutrons and protons with temperature are very similar to each other
since the neutron and proton number are the same.
From zero temperature to high temperature, the intruder level $9/2^+[404]$ from the $1g_{9/2}$ orbital, which drives the nucleus oblate, gradually goes above the Fermi surface.
The occupation on this intruder level becomes less and less with increasing temperature. Correspondingly, the oblate deformation becomes smaller and smaller, and eventually goes to zero at $T\sim2.1$ MeV, where the s.p. levels of the same angular momentum quantum number become degenerate and the spherical s.p. levels are formed.
The large subshell gaps developed at $N(Z)$=36 contribute for stabilizing the minimum states
before the continuous deformation change at 0.9 MeV.

\begin{center}
\includegraphics[scale=0.65]{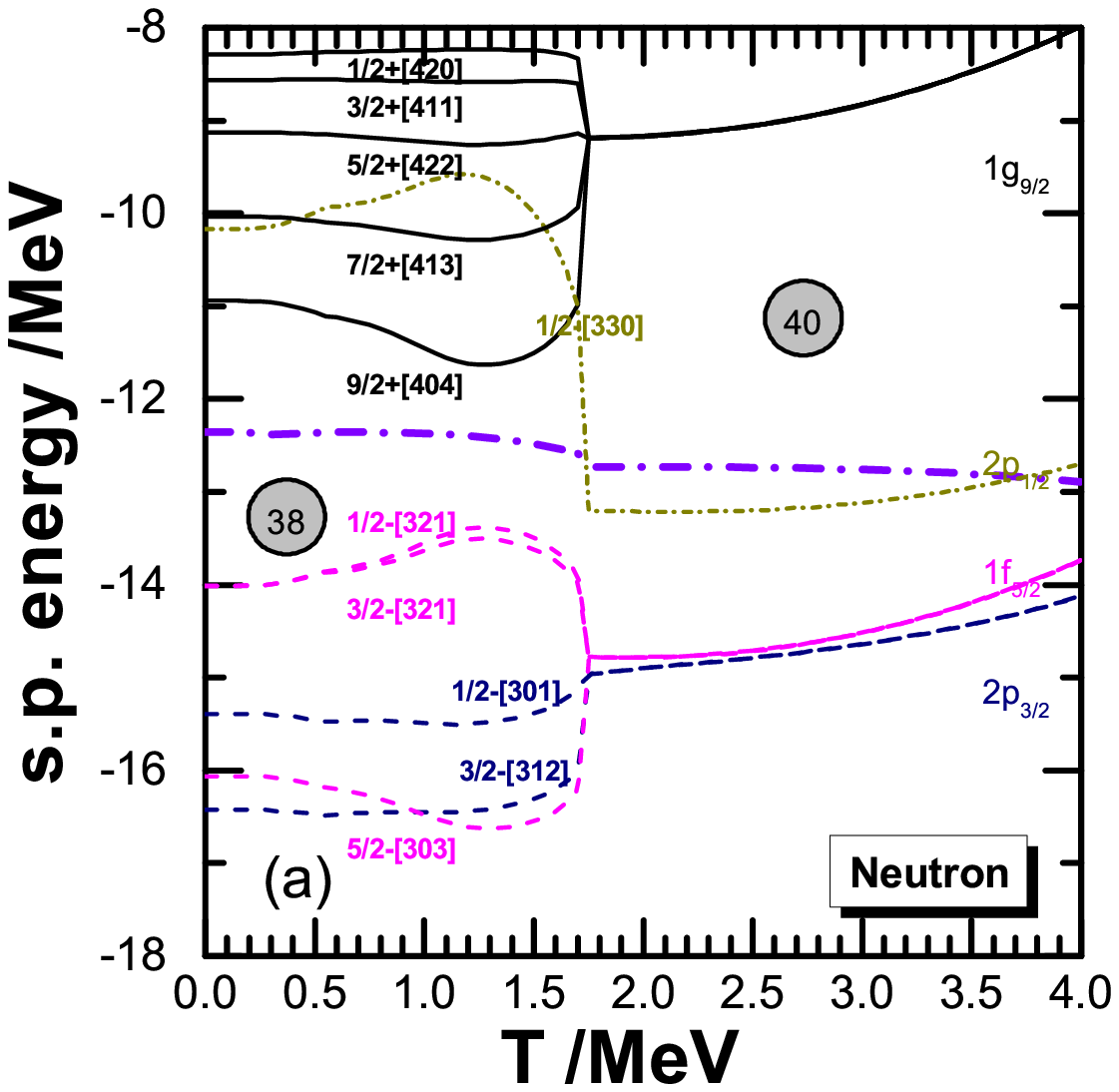}%
\includegraphics[scale=0.65]{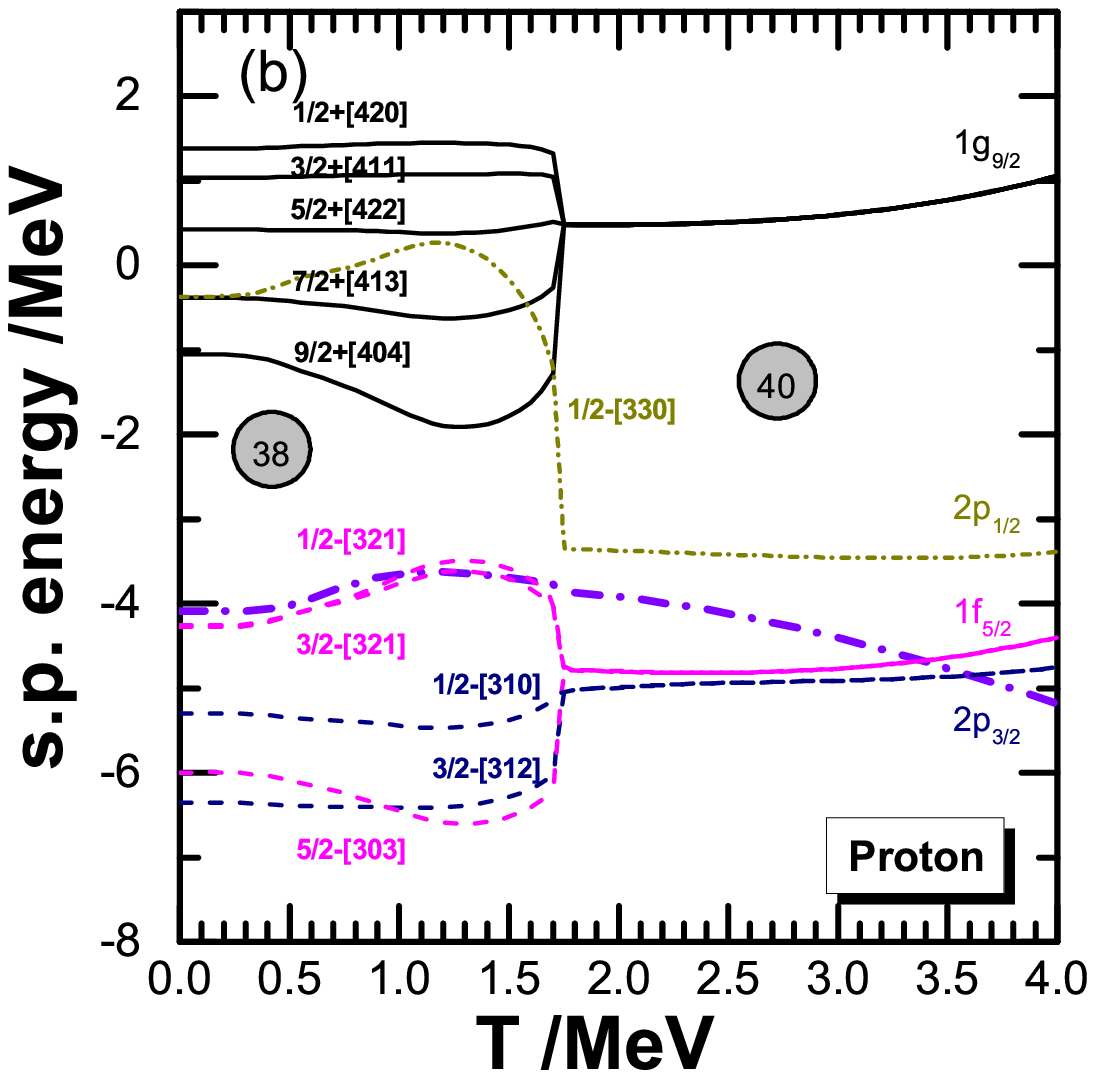}\\
\figcaption{\label{Kr74sp}
Same as Fig. ~\ref{Kr72sp}, but for $^{74}$Kr.
}
\end{center}

In Fig.~\ref{Kr74sp}, the s.p. levels of $^{74}$Kr behaves differently from those of $^{72}$Kr.
Since the global minimum locates near $\beta_2=-0.14$ before temperature 1.7 MeV, and
locates near spherical after that temperature,
this evolution is basically a direct jump from oblate to prolate at $T=1.7$ MeV.
It can be seen that the large subshell gap developed at $N=$38
contributes for stabilizing the minimum states at lower temperatures.

\begin{center}
\includegraphics[scale=0.3]{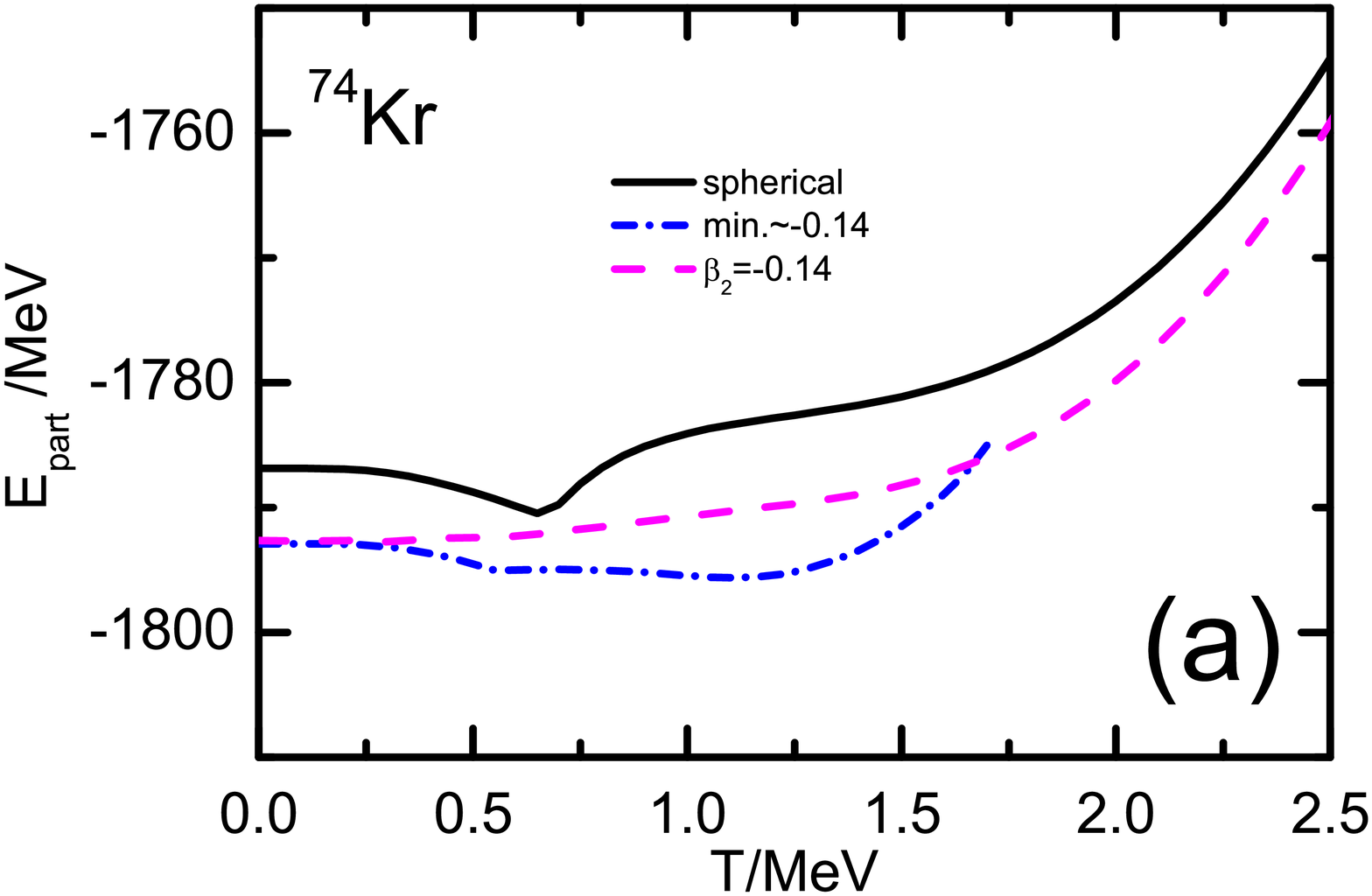}%
\includegraphics[scale=0.3]{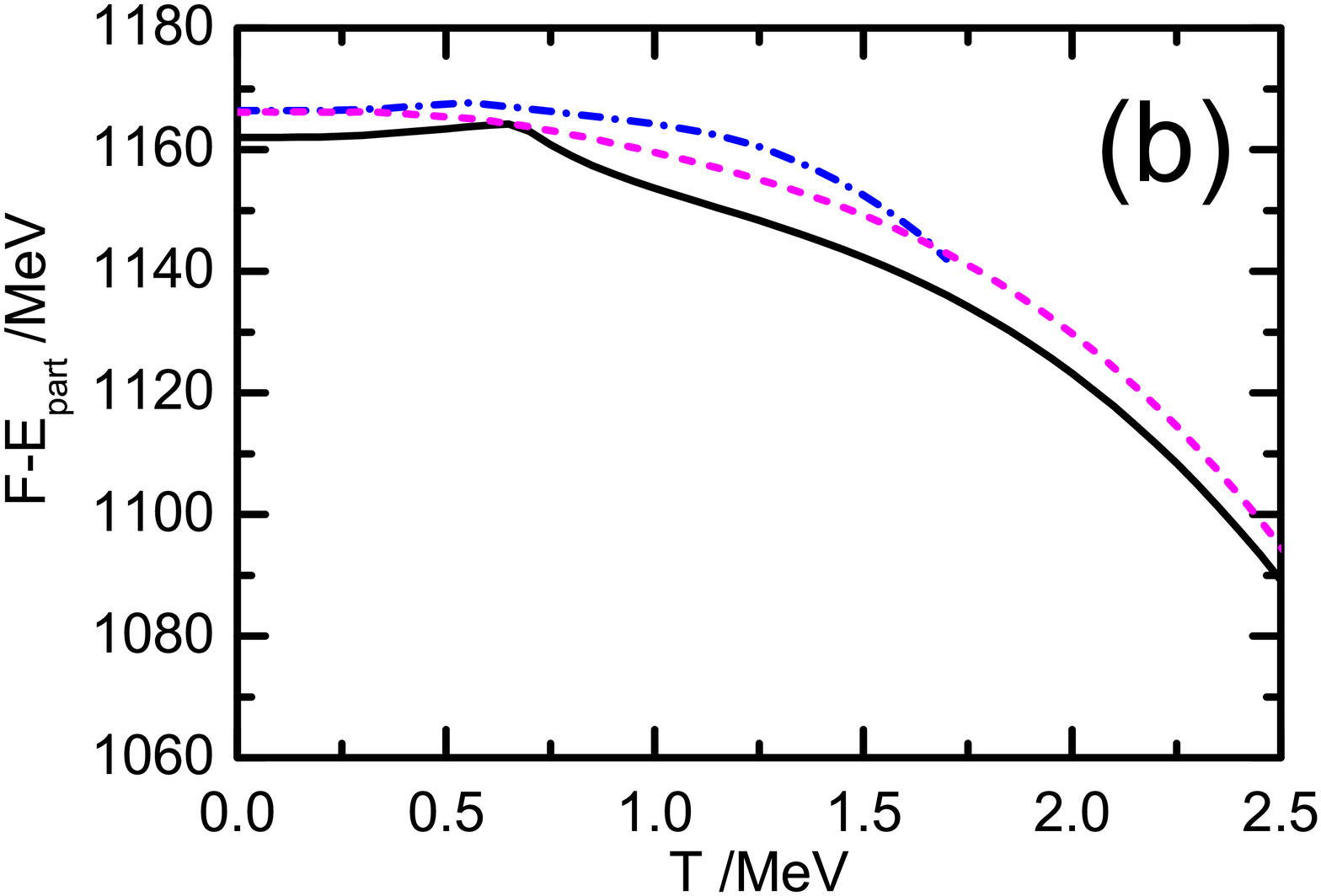}\\
\includegraphics[scale=0.3]{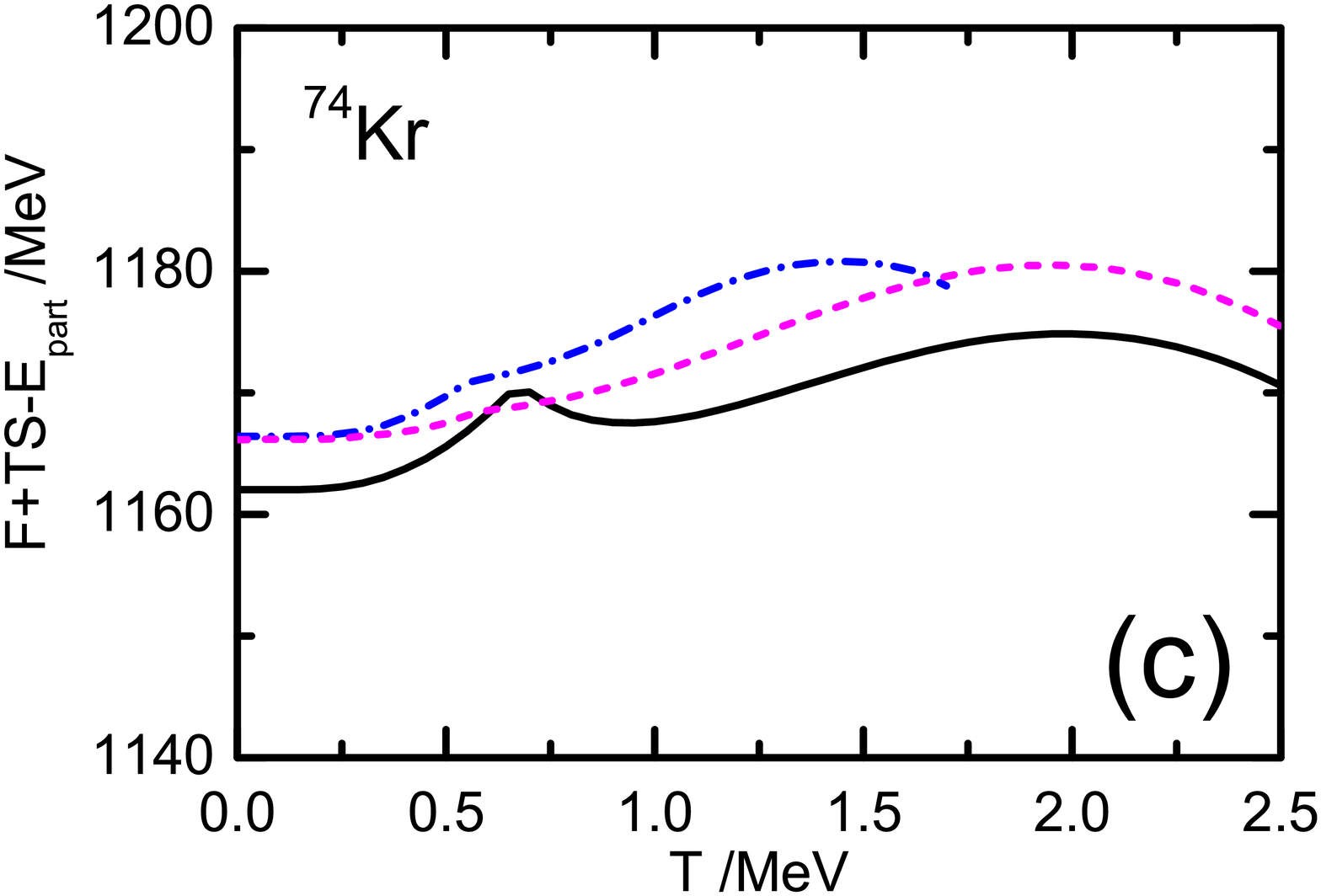}
\includegraphics[scale=0.3]{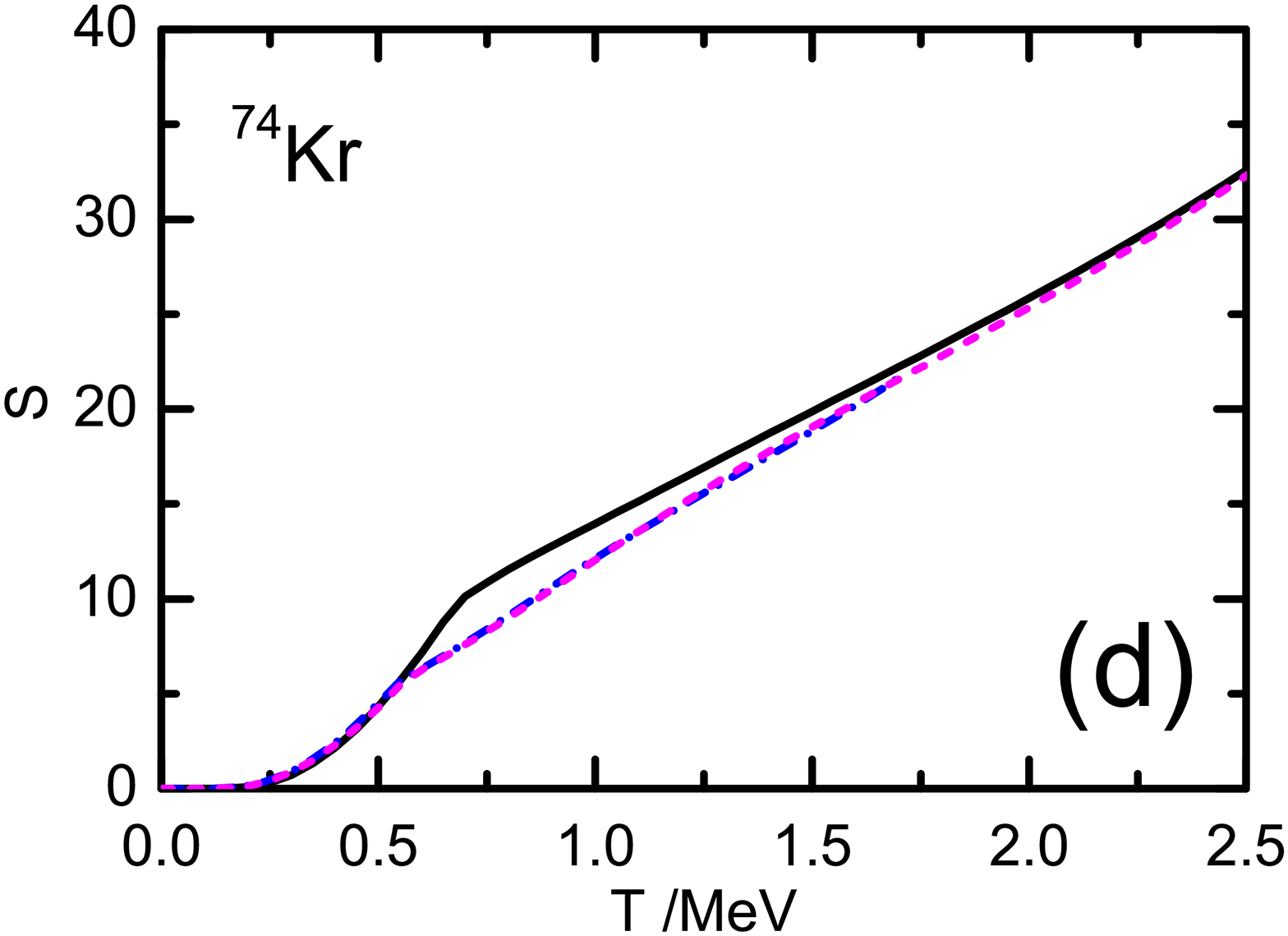}\\
\includegraphics[scale=0.3]{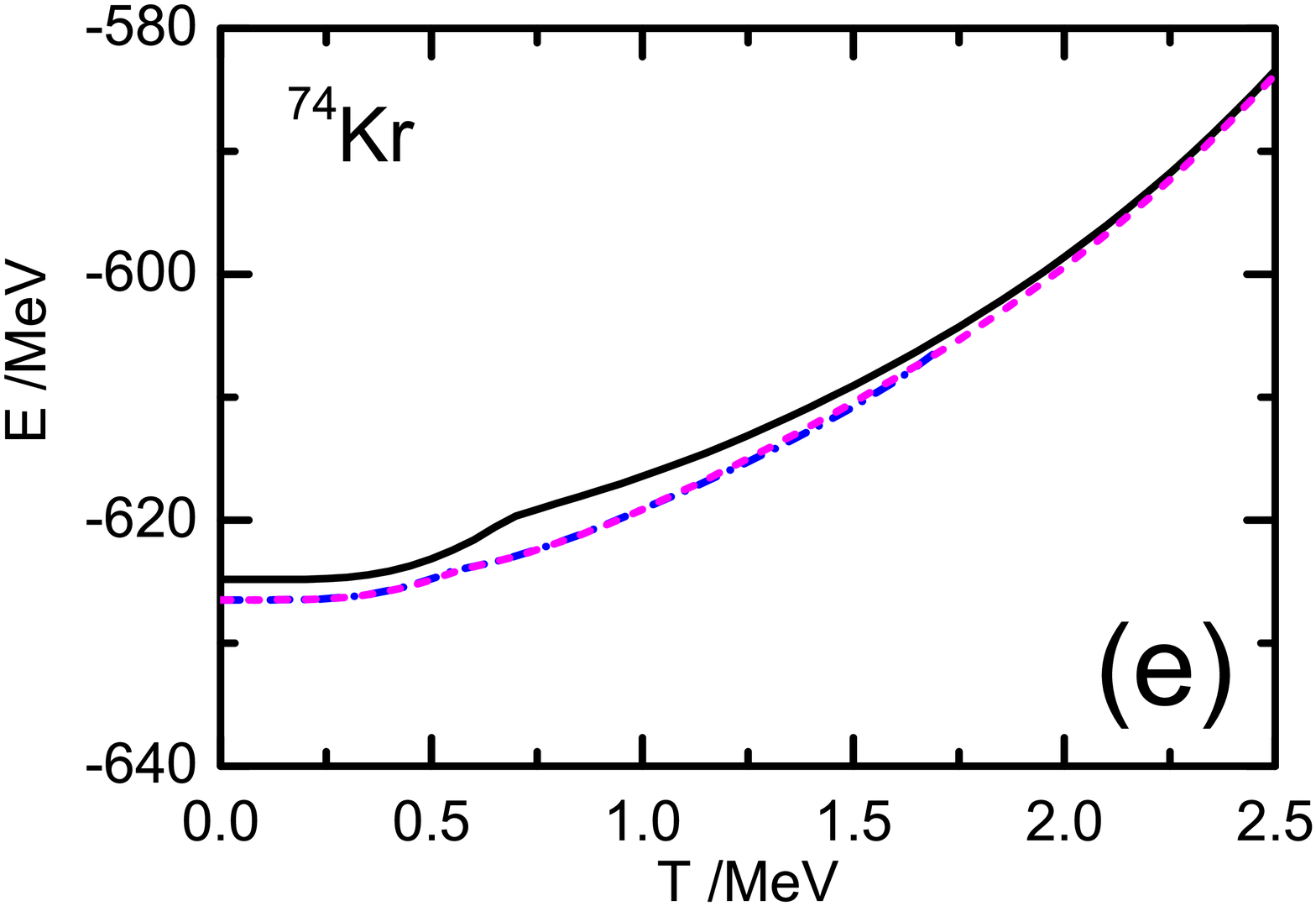}
\includegraphics[scale=0.3]{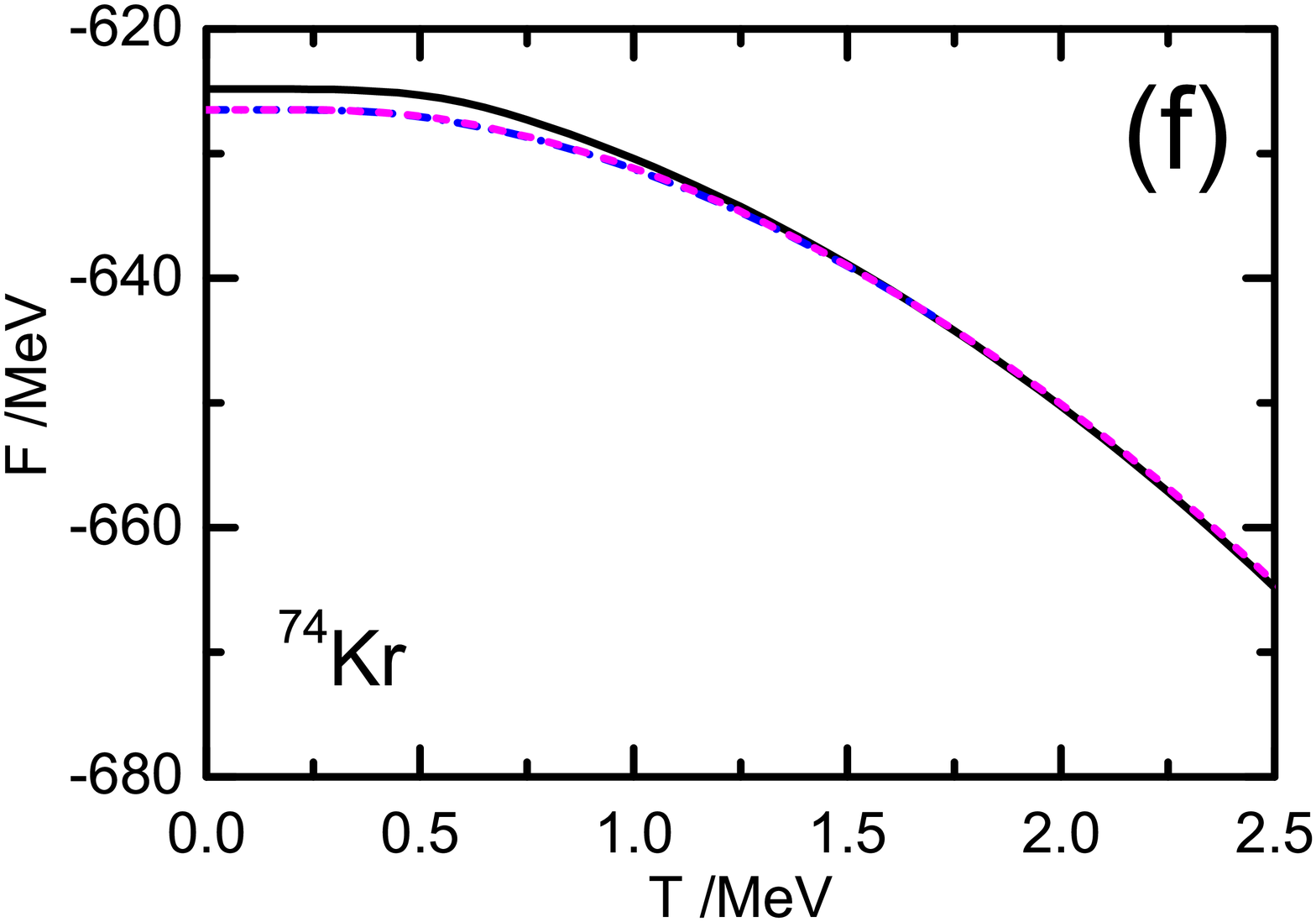}\\
\figcaption{\label{Kr74E}
The particle energies (a), energy difference between the free energy and particle energy (b),
energy difference between the internal binding energy and particle energy (c), the entropy (d),
the internal energy (e), and the free energy (f) for global minimum near $\beta_2 \sim -0.14$, state at exactly $\beta_2=-0.14$, and the spherical state, as functions of temperature (in MeV) for $^{74}$Kr, obtained by the constrained CDFT+BCS calculations using PC-PK1 energy density functional.
}
\end{center}

In order to understand the changes of dominant shape
with temperature in $^{74}$Kr, we study the evolution of
the energies composing the free energy with rising temperatures in Fig.~\ref{Kr74E}.
Firstly we decompose the free energy into two parts, the particle energy $E_{\rm part}$ and $F-E_{\rm part}$.
The particle energy $E_{\rm part}$ is the total sum of the occupied single-particle energies, which reflects the shell effects of the Nilsson diagram at the zero temperature case, as well as the decreasing shell effects by the changing thermal occupation probabilities at finite temperature cases.
At zero temperature, the particle energy usually get minimized at the deformation where a shell gap above the last occupied nucleons appears in the Nilsson diagram.
So we can see that indeed the particle energy for the minimum at $\beta_2\sim$-0.14,
where there is a big shell gap at $N$=38 in Fig.~\ref{Kr74sp}, is smaller than that at spherical from Fig.~\ref{Kr74E}(a).
Furthermore, the second part $F-E_{\rm part}$ can be decomposed into two parts,
the field energy $F+TS-E_{\rm part}=E-E_{\rm part}$ and the product of the temperature and the entropy $-TS$.
The field energy mainly represents the mean field potential energy,
namely the contributions from the isoscalar-scalar, isoscalar-vector, isovector-vector, and electromagnetic fields,
which is explained in Sec.~\ref{sec2}\normalfont.
The field energy normally prefers the spherical shape.
So it is expected to observe in Fig.~\ref{Kr74E}(c) that
the spherical shape has a smaller field energy than the deformed shape.
From the curves of spherical shape in Fig.~\ref{Kr74E}(a)-(e), we notice that there is a kink at $T \sim$ 0.6 MeV, which actually corresponds to the disappearance of the pairing gap at spherical shape with increasing temperature.
This disappearance of the pairing gap at spherical shape actually boosts the entropy in Fig.~\ref{Kr74E}(d) at $T\sim 0.6$ MeV.
As a result, it lowers the $F-E_{\rm part}$  in Fig.~\ref{Kr74E}(b), even if the field energy of the spherical shape in Fig. ~\ref{Kr74E}(c) increases at the kink. If we compare Fig.~\ref{Kr74E}(e) and Fig.~\ref{Kr74E}(f), we can notice that without the inclusion of entropy, the spherical shape becomes lower in energy than the oblate shape
at a higher temperature $T\sim$ 2.5MeV.
The kink behavior of the entropy at pairing transition temperature $T\sim 0.6$ MeV for the spherical shape substantially affects the transition temperature $T\sim 1.7$ MeV from oblate to spherical shape .

\section {Summary}
In summary,
the finite-temperature axially deformed CDFT + BCS theory based on the relativistic point-coupling density functional
is developed in this paper, and applied to the shape evolution study of $^{72,74}$Kr with temperature.
For $^{72}$Kr, with temperature increasing,
the nucleus changes from oblate to a spherical shape at $T\sim2.1$ MeV
with a relatively quick deformation change at $T\sim0.9$ MeV.
For $^{74}$Kr, its global minimum locates at $\beta_2 =-0.14$
and abruptly change to spherical at $T\sim 1.7$ MeV.
The proton pairing transition occurs at $T=0.6$MeV following the rule $T_c =0.6 \Delta_p(0)$
due to stable deformation for the global minimum with rising temperature.
The signatures of the above pairing transition or shape changes can
be found in the curve of the specific heat.
The single-particle level evolutions with the temperature as well as the deformation are presented.
The large subshell gap developed at $N(Z)$=36 or 38 contributes for stabilizing the minimum states
for low temperatures for $^{72}$Kr or $^{74}$Kr respectively.
As an initial work on the investigation of shape evolution with temperature for these complicated nuclei, our study provides a qualitative understanding on the evolution picture. However, to quantitatively describe this phenomena, one needs to go beyond mean-field approximation, for example, by developing the finite-temperature GCM theory. The corresponding work is envisaged.

\acknowledgments{
We sincerely express our gratitude to Jiang Ming Yao for helpful discussions.
The theoretical calculation was supported by the nuclear data storage system in Zhengzhou University.}

\vspace{-1mm}
\centerline{\rule{80mm}{0.1pt}}
\vspace{2mm}

\begin{multicols}{2}

\end{multicols}

\clearpage

\end{document}